\documentclass[aps,english,nofootinbib,notitlepage,showpacs,longbibliography]{revtex4-1}
\usepackage{amstext,amsmath,amssymb,amsfonts,bbm}
\usepackage[latin1]{inputenc}
\usepackage{graphicx}
\usepackage{epstopdf}
\usepackage{amsthm}
\usepackage{tocvsec2}
\usepackage{enumerate}
\usepackage{subfigure}
\usepackage{color}

\usepackage{multirow} \usepackage{rotating}
\usepackage[colorlinks,citecolor=blue,linkcolor=blue,urlcolor=blue]{hyperref}
\usepackage{ae,aecompl}

\usepackage{babel}
\usepackage{float}

\usepackage{amssymb}
\usepackage{graphicx}
\usepackage{esint}

\def\beq{\begin{equation}}
\def\be{\begin{equation}}
\def\ee{\end{equation}}
\def\bes{\begin{eqnarray}}
\def\ees{\end{eqnarray}}



\begin{document}

\title{Contracted Bianchi Identity and Angle Relation on $n$-dimensional
Simplicial Complex of Regge Calculus}

\author{Seramika Ariwahjoedi$^{1}$, Freddy P. Zen$^{1,2}$\vspace{1mm}
}

\affiliation{$^{1}$Theoretical Physics Laboratory, THEPI Division, Institut Teknologi
Bandung, Jl. Ganesha 10 Bandung 40132, West Java, Indonesia.\\
$^{2}$Indonesia Center for Theoretical and Mathematical Physics (ICTMP),
Indonesia.}

\begin{abstract}
\noindent In this article, we prove the theorems concerning the trace
relation of SO(3), SU(2), and SO(n) which are representation of SO(3)
and SU(2). An interesting fact we found is the trace relation of SU(2)
gives the spherical law of cosine which in turns is a dihedral angle
relation, a constraint that must be satisfied by closed Euclidean
simplices. Moreover, we applied our results on general group elements
to holonomies on the simplicial complex of Regge Calculus, which is
the main motivation of this article. Here, we found that: (1) in 4-dimensional
Euclidean Regge Gravity, all the holonomy circling a single hinge
are simple rotations, and (2) the dihedral angle relation represents
the 'contracted' Bianchi identity for a simplicial complex.
\end{abstract}
\maketitle

\section{Introduction}

The discrete attempt on gravity was first introduced by Tullio Regge,
written in the second order formulation where the generalized coordinate
is the spatial metric \cite{Regge1,Regge2}. This discrete formulation,
theoretically works for any dimension. In particular, for 3-dimensional
gravity, the model is known as the Ponzano-Regge model \cite{Ponzano,Barret1}.
Another important development on general relativity was carried in
\cite{Cartan1,Cartan2}, where it is written in the form closer to
gauge theory, usually known as the first order formulation. This is
followed by the introduction of new variables in \cite{Ashtekar},
which is important for the canonical quantization of gravity. The
discrete version of the first order formulation of gravity immediately
becomes an interest in the quantum gravity community, with the first
development carried in \cite{Barret2}. This is followed by the Barret-Crane
model for 4-dimensional Lorentzian gravity \cite{Crane1,Crane2}.
Some corrections on the Barret-Crane model gives the EPRL-FK (or spinfoam)
model \cite{Engle,Baez1,Perez}, which could be derived from discrete
BF theory. All these model use a simplicial complex to described spacetime,
with the difference among them being the variations on action integral.

We are interested in discrete gauge theory of gravity, particularly,
in the description of the possible simplicial complex of Regge gravity.
It is well-known that discrete manifold in Regge Calculus is a special
case of Riemannian manifold, in the sense that their Riemann tensor
must be in the form of $R_{\mu\nu\beta}^{\alpha}\sim\hat{\omega}_{\:\,\beta}^{\alpha}\hat{\omega}_{\mu\nu}$
\cite{Sato,Carlo1}. With this restriction, it is natural to ask what
would be the holonomy group for the discrete manifold in Regge Calculus.
The importance of the holonomy in the theory is crucial if one consider
the possibility of the discreteness of space in the Planck scale,
as predicted by loop quantum gravity and other non-perturbative theories
of gravity \cite{Carlo2,Ashtekar2,Ashtekar3}. 

As a first step to solve the problem, we consider several important
aspects of the theory: (1) The rotation group in 3-dimension, SO(3)
and its double-cover, SU(2). These groups become our interest because
it is the natural gauge group of the 3-dimensional spatial part of
the spacetime bundle. In this article, we will show that the importance
of these groups for a simplicial complex, particularly SU(2), is not
only restricted to dimension (3+1). The second is (2) the 'dihedral'
angle relation of a simplex. One has the fact that the $d$ and $(d-1)$-dimensional
angles in a closed $(n\geq d)$-simplex satisfy this relation as a
constraint \cite{Simone}. In three dimension, the relation is automatically
satisfied by three bivectors meeting in a point, but this is not the
case in four and larger dimension. In fact, it is shown in this article
that one of the gauge groups in point (1) will give rise to the dihedral
angle relation as its trace relation (or contracted Bianchi Identity
for holonomy). Thus, one needs to take this relation into account.
The last aspect related to the angle relation is (3) the simplicity
of a bivector, and moreover the simplicity of the rotations. Similar
with the angle relation, this is automatically satisfied in 3D but
not in larger dimension. We will show that these three properties
are related to one another and are important for the construction
of a simplicial complex in $n$-dimension. Specifically, the gauge
group in $n$-simplicial complex of Regge Calculus is SO(n), but certain
restrictions are needed, in order to described a theory of discrete
gauge gravity. 

The organization of the article is the following: Section II consists
a discussion of the asymptotics of second Bianchi Identity, where
the discreteness of space is described by holonomies o hinges: elements
of Lie group attached as a variables on the loops. In the next two
sections, we study a special case of gauge group SO(n) in general,
without refering to the loops. Specifically, Section III consists
the proof of theorems concerning the trace relation of elements of
SO(3), SU(2), and SO(n) which are representation of SO(3) and SU(2).
Here, we found that the trace relation of SU(2) gives the spherical
law of cosine which in turns, is equivalent with the angle relation.
The application of some lemmas in Section III to the 4-dimensional
case gives a classification of rotations in 4D Euclidean space, this
is discussed in Section IV. In Section V, we applied our theorems
on general group elements to holonomies on the simplicial complex
of Regge Calculus, which is the main motivation of this article. Here,
we found that: (1) in 4-dimensional Euclidean Regge Gravity, all the
holonomy circling a single hinge of a simplicial complex are simple
rotations, and (2) the dihedral angle relation represents the 'contracted'
Bianchi identity for a simplicial complex. Finally at the last section,
we discuss the relevance of our findings to the established theory
of discrete gauge gravity.

\section{Second Bianchi Identity}

\subsection{Second Bianchi Identity in a point and finite loops}

Given a fibre bundle $\mathcal{E}$ diffeomorphic to a standard fibre
bundle $\mathcal{M}\times\mathcal{F}$, with $\boldsymbol{A}$ is
the connection on $\mathcal{E},$ the second Bianchi Identity is the
condition that must be satisfied by the curvature 2-form $\boldsymbol{F}=d_{D}\boldsymbol{A}$:
\begin{equation}
d_{D}\boldsymbol{F}=D_{\mu}F_{\nu\lambda}dx^{\mu}\wedge dx^{\nu}\wedge dx^{\lambda}=\frac{1}{3}\left(D_{\mu}F_{\nu\lambda}+D_{\nu}F_{\lambda\mu}+D_{\lambda}F_{\mu\nu}\right)dx^{\mu}\otimes dx^{\nu}\otimes dx^{\lambda}=0.\label{eq:prev}
\end{equation}
The geometrical interpretation of second Bianchi Identity is clear
in the finite setting. 
\begin{figure}
\begin{centering}
\includegraphics[scale=0.6]{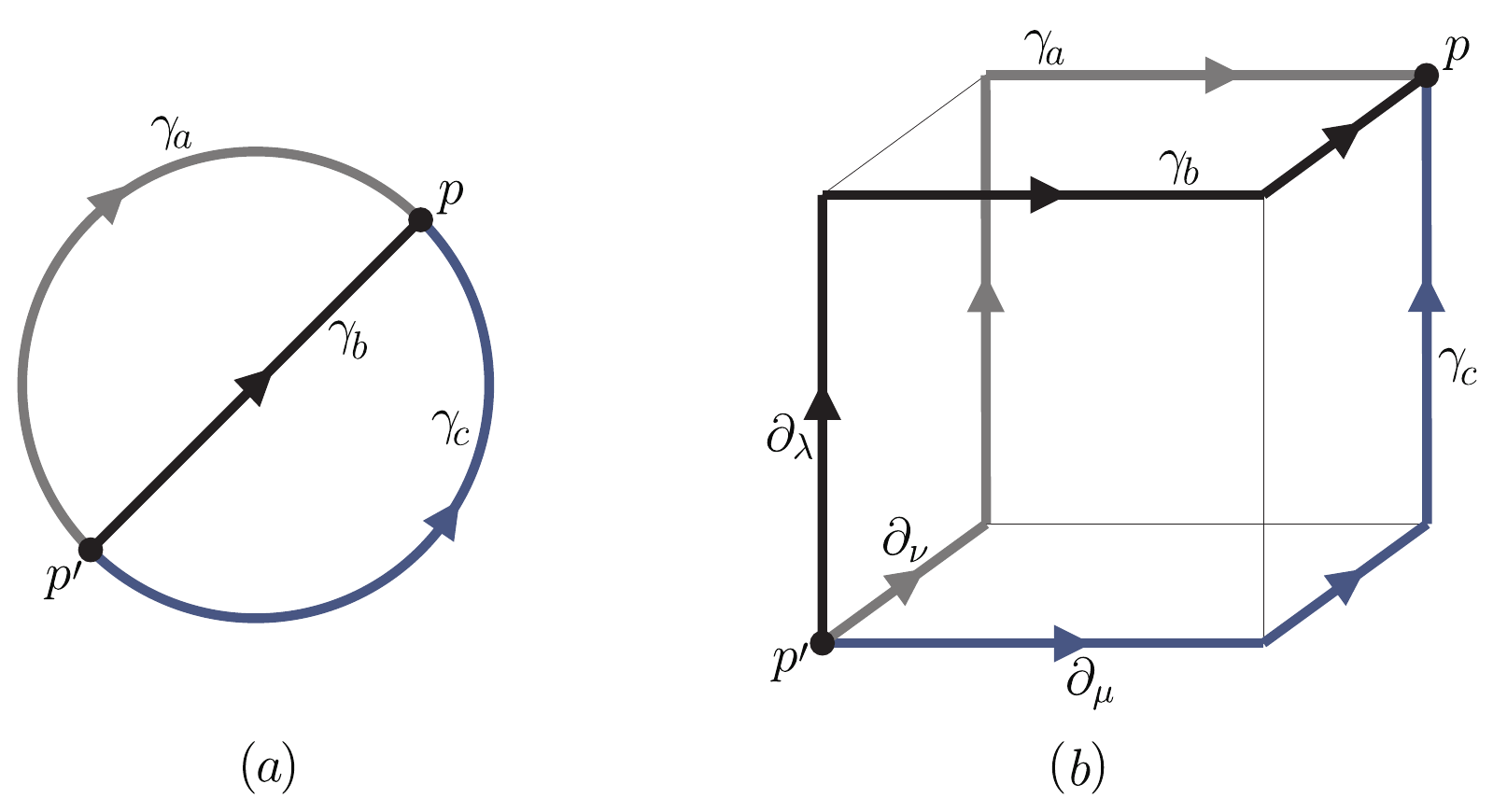}
\par\end{centering}

\caption{(a). Given the loops on the \textit{theta} graph configuration as
follows: $\gamma_{1}=\gamma_{a}\gamma_{b}^{-1},$ $\gamma_{2}=\gamma_{b}\gamma_{c}^{-1},$
and $\gamma_{3}=\gamma_{c}\gamma_{a}^{-1},$ the holonomies on loops
$\gamma_{i}$, $i=1,2,3$ satisfy the second Bianchi Identity (\ref{eq:bifinite}).
(b). Approximation of the theta configuration (a) by the square loops.}
\end{figure}
 Given three loops $\gamma_{i}$ on $\mathcal{M}$ such that they
meet together on two points as in FIG. 1a, the finite second Bianchi
Identity is a product of three holonomies $U_{\gamma_{i}}=U_{\gamma_{i}}\left(\boldsymbol{A},p\right)$:
\begin{equation}
U_{\gamma_{1}}U_{\gamma_{2}}U_{\gamma_{3}}=1,\label{eq:bifinite}
\end{equation}
with $p\in\mathcal{M}$ is the origin of the loop. Each holonomy is
a solution to a parallel transport along the loop, $\frac{dU_{\gamma_{i}}}{d\tau}=0$:
\begin{equation}
U_{\gamma_{i}}=\hat{P}\exp\oint_{\gamma_{i}=\partial S_{i}}\boldsymbol{A}=\hat{P}\exp\int_{S_{i}}d\boldsymbol{A}=\hat{P}\exp\int_{S_{i}}\boldsymbol{F},\label{eq:aa}
\end{equation}
where the last equality is obtained by applying Stokes theorem on
a closed loop $\gamma_{i}$ as the boundary of a surface $S_{i}$.
A holonomy, by definition (\ref{eq:aa}), is an element of a gauge
group $\mathcal{G}$, which is attached as a variable on a loop $\gamma$
$\subset$ $\mathcal{M}$. In analog with the relation between a Lie
group and its algebra, it comes from the 'exponential map' (\ref{eq:aa})
of the connection $\boldsymbol{A}$, which is a Lie algebra valued
1-form on $T_{p}^{*}\mathcal{M}.$ 

For a case where loops $\gamma_{i}$ are infinitesimally small, they
can be approximated to a square loop with length $\varepsilon^{2}$
as in FIG. 1b. Expanding (\ref{eq:bifinite}) in terms of $\varepsilon$
and taking only the first non-zeroth order, one obtains (\ref{eq:prev})
such that it can be thought as a relation defined on an infinitesimal
loop. 

In the rest of the article, we will refer the second Bianchi Identity
simply as the Bianchi Identity.

\subsection{Asymptotics of Bianchi Identity}

With the curvature and the scale $\varepsilon$ as parameters, one
could obtain asymptotics of the Bianchi Identity. Thus, one has four
conditions, where the first is the case with no approximation, namely,
the Bianchi Identity defined on a finite loop (finite scale $\varepsilon$)
and finite (non-zero) curvature. The holonomies are exactly (\ref{eq:aa}),
and the Bianchi Identity is (\ref{eq:bifinite}).

The second case is the nearly-flat case with finite loop and infinitesimally
small curvature. For small $\boldsymbol{F}$, the holonomy can be
approximated as follows:
\begin{equation}
U_{S_{i}}\approx I+\intop_{S_{i}}\boldsymbol{F},\label{eq:flat1}
\end{equation}
where it is labeled by the surface $S_{i}$ enclosed by loop $\gamma_{i}$,
and $I$ is the identity. The Bianchi Identity becomes a closure condition:
\begin{equation}
\intop_{S_{i}}\boldsymbol{F}+\intop_{S_{j}}\boldsymbol{F}+\intop_{S_{k}}\boldsymbol{F}\approx0,\label{eq:flat2}
\end{equation}
where the curvature $\boldsymbol{F}$ is smeared on each finite surface.

The next case is the nearly-continuous case with infinitesimal scale
and finite curvature. For the reason explained in Section II A, the
holonomies could be approximated into:
\[
U_{S_{\mu}}\approx I+\frac{\varepsilon^{2}}{2}\left(F_{\mu\nu}-F_{\lambda\mu}\right)+\frac{\varepsilon^{3}}{3!}D_{\mu}F_{\nu\lambda},
\]
and the Bianchi Identity (\ref{eq:prev}) comes from the third order
of $\varepsilon$.

The last case is the nearly-flat and nearly-continuous case. The holonomy
is approximated into:
\[
U_{S_{i}}\approx I+\varepsilon^{2}\boldsymbol{F}\left(l_{j},l_{k}\right),
\]
with $S_{i}\cong l_{j}\wedge l_{k}$, and the corresponding Bianchi
Identity comes from the second order of $\varepsilon$:
\[
\boldsymbol{F}\left(l_{j},l_{k}\right)+\boldsymbol{F}\left(l_{k},l_{i}\right)+\boldsymbol{F}\left(l_{i},l_{j}\right)\approx0.
\]

A recent growing interest in discrete gauge theory of gravity and
quantum gravity is the idea that space is fundamentally discrete in
the Planck scale \cite{Carlo2,Ashtekar2,Ashtekar3}, and that curvature
is quantized such that non-trivial loops could not be shrunk to a
point \cite{Thiemann}. Therefore the relevant case of holonomies
on these non-contractible loops are the finite loop cases, namely
(\ref{eq:bifinite})-(\ref{eq:aa}) and (\ref{eq:flat1})-(\ref{eq:flat2}).
Up to the rest of the article, we will only consider these cases.

\section{Contracted Bianchi Identity and Dihedral Angle Formula}

Following the classification of manifold by their holonomy groups
in \cite{Berger}, the holonomy group of a general, oriented, $n$-dimensional
Riemannian manifold is (a subgroup of) the special orthogonal group
SO(n). Some special condition on SO(n) will give rise to special properties
on the Riemannian manifold. We will show that the holonomy group of
an oriented, $n$-dimensional simplicial complex in Euclidean Regge
Calculus needs to be SO(n) plus some certain condition. In order to
show this, in the following section we study the holonomy as an element
of a Lie Group, which could be defined independently from the loop
on the base manifold. The theorem we obtain in this section will give
important geometrical meaning if applied on loops, which are discussed
in the last section.

\subsection{The case of SO(3) as gauge group}

In gauge theory, it is important to obtain the gauge-invariant quantities.
One of them is the trace of holonomy, or in physics term, the Wilson
loop:
\[
\chi=\textrm{tr }U_{\gamma_{i}}=\textrm{tr }\hat{P}\exp\oint_{\gamma_{i}}\boldsymbol{A}.
\]
If $U_{i}$ is a matrix representation of an element of a group, the
trace is the characteristics polynomials of the matrix. The physical
importance of the Wilson loop is the Lagrangian (or action integral)
in Chern-Simon theory.

A gauge-invariant quantity which become our interest is the \textit{trace
relation} of a gauge group as follows:
\begin{equation}
\textrm{tr }U_{3}^{-1}=\textrm{tr }U_{1}U_{2},\label{eq:3.1}
\end{equation}
which is clearly equivalent with the contracted Bianchi Identity.
Our pre-result in the previous paper \cite{paper} is: 

\textbf{Theorem 3.1 (Trace Relation of Gauge Group SO(3))}

\textit{Given elements of group $U_{i}\in SO(3)$, the trace relation
or the contracted Bianchi Identity in the form of (\ref{eq:3.1}),
gives the relation as follows:}
\begin{equation}
\cos\frac{\theta_{3}}{2}=\pm\left(\cos\frac{\theta_{1}}{2}\cos\frac{\theta_{2}}{2}-\cos\phi_{12}\sin\frac{\theta_{1}}{2}\sin\frac{\theta_{2}}{2}\right),\label{eq:3.2}
\end{equation}
\textit{with $\theta_{i}$ are the rotation angles of $U_{i}$} \textit{and
$\phi_{12}$ is the angle between plane of rotation of $U_{1}$ and
$U_{2}$.}

\textbf{Proof.} The proof for this theorem is direct and straightforward,
realizing that any element $U_{i}\in SO(3)$ can always be written
as:\textit{
\begin{equation}
U_{i}=I_{3\times3}+\hat{J}_{i}\sin\theta_{i}+\hat{J}_{i}^{2}\left(1-\cos\theta_{i}\right),\qquad J_{i}\in\mathfrak{so(3)}.\label{eq:3.3}
\end{equation}
}Using the following relations for plane of rotations $J_{i}=\vec{J}_{i}\cdot l\in\mathfrak{so(3)},$
with $\vec{J}_{i}\in\mathbb{R}^{3}$ and $l$ are generator of $\mathfrak{so(3)}$:
\begin{align*}
\textrm{tr}\left(J_{i}J_{j}\right)= & -2\left\langle \vec{J}_{i},\vec{J}_{j}\right\rangle \\
\textrm{tr}\left(J_{i}J_{j}J_{k}\right)= & \vec{J}_{i}\cdot\left(\vec{J}_{j}\times\vec{J}_{k}\right)\\
\textrm{tr}\left(J_{i}J_{j}J_{k}J_{l}\right)= & \left\langle \vec{J}_{i},\vec{J}_{l}\right\rangle \left\langle \vec{J}_{j},\vec{J}_{k}\right\rangle +\left\langle \vec{J}_{i},\vec{J}_{j}\right\rangle \left\langle \vec{J}_{k},\vec{J}_{l}\right\rangle ,
\end{align*}
together with the half angle formula, one could obtain dihedral relation
(\ref{eq:3.3}). 

The aim of this article is to show that special restrictions on the
gauge group SO(n) in any dimension will give rise to the same dihedral
angle relation. This can be thought as an immersion of SO(3) on SO(n).
It is clear that the image of the 'immersion' map $\rho_{n}$: 
\[
\rho_{n}:SO\left(3\right)\rightarrow SO(n),
\]
is a representation of SO(3) in $n$-dimension. The map induces a
representation of the corresponding Lie algebra as follows:
\[
d\rho_{n}:\mathfrak{so\left(3\right)}\rightarrow\mathfrak{so\left(n\right)},
\]
The proof for this is simply to show that $\rho_{n}$ is a group homomorphism. 

Before arriving at the result, the followings are supporting definitions
and lemmas used to obtain the theorems. Let $\mathbb{R}^{n}\times\mathbb{R}^{n}$
be a space of matrices. Using terminologies in $\mathbb{R}^{n}$,
the coordinate basis in $\mathbb{R}^{n}\times\mathbb{R}^{n}$ is \textit{$\left\{ dx^{i}\otimes dx^{j}\right\} ,$
}for\textit{ $i,j=1,..,n$}, such that any element of the space can
be written as their linear combination: $u=u_{ij}dx^{i}\otimes dx^{j}.$
The matrix space $\mathbb{R}^{n}\times\mathbb{R}^{n}$ could be equipped
with the Frobenius inner product defined as follows:
\begin{eqnarray*}
\left\langle \:,\:\right\rangle :\left(\mathbb{R}^{n}\times\mathbb{R}^{n}\right)\times\left(\mathbb{R}^{n}\times\mathbb{R}^{n}\right) & \rightarrow & \mathbb{R}\\
\left(u,v\right) & \mapsto & \left\langle u,v\right\rangle ,
\end{eqnarray*}
where the operation is given by:
\begin{eqnarray}
\left\langle u,v\right\rangle  & = & u_{ij}v_{kl}\delta^{ik}\delta^{jl},\label{eq:3.16-1-1}
\end{eqnarray}
with $\delta^{ik}\delta^{jl}$ comes from the orthonormality condition
of the coordinate basis: 
\begin{equation}
\delta^{ik}\delta^{jl}=\left\langle dx^{i}\otimes dx^{k},dx^{j}\otimes dx^{l}\right\rangle .\label{eq:3aa-1-1}
\end{equation}

Other properties of matrices which are important particularly in this
articles are the following: The trace of matrix $\omega$ is defined
as $\textrm{tr }\omega=\omega_{ij}\delta^{ij},$ matrix multiplication
of two matrices $u$ and $v$ is defined as $uv=u_{ij}v_{kl}\delta^{jk}\left(dx^{i}\otimes dx^{l}\right),$
and transpose of a matrix $\omega$ is defined as $\omega^{T}=\omega_{ji}dx^{i}\otimes dx^{j}.$
With these definitions, the Frobenius inner product (\ref{eq:3.16-1-1})
could be written as $\left\langle u,v\right\rangle =\textrm{tr}\left(u^{T}v\right).$

Let $\left\{ \hat{e}^{ij}\right\} $ be a non-coordinate basis (not
necessarily orthonormal) of $\mathbb{R}^{n}\times\mathbb{R}^{n}$
given by the Gramm-Schmidt procedure. $\left\{ \hat{e}^{ij}\right\} $
can always be written as:
\begin{equation}
\hat{e}^{ij}=e_{\;\;\, kl}^{ij}dx^{k}\otimes dx^{l},\label{eq:ais}
\end{equation}
where $e_{\;\;\, kl}^{ij}$ is a rank four tensor. Using the fact
that $\mathbb{R}^{n}\times\mathbb{R}^{n}$ is isomorphic with $\mathbb{R}^{n^{2}}$,
$e_{\;\;\, kl}^{ij}$ needs to be invertible on the pair of indices
$(ij)$ and $(kl)$, namely the inverse $\left(e^{-1}\right)_{\;\;\,\left(kl\right)}^{\left(ij\right)}=\left(e_{\;\;\,\left(kl\right)}^{\left(ij\right)}\right)^{-1}$
exists. Two sets of non-coordinate basis $\left\{ \hat{e}^{ij}\right\} $
and $\left\{ \hat{e}'^{ij}\right\} $ are similar up to a general
transformation $\Omega$ as follows:
\begin{equation}
\hat{e}'{}^{ij}=\sum_{k,l}\Omega^{\left(ij\right)\left(kl\right)}\hat{e}^{kl},\label{eq:huehj}
\end{equation}
Since $\left\{ \hat{e}^{ij}\right\} $ and $\left\{ \hat{e}'^{ij}\right\} $
are basis of $\mathbb{R}^{n}\times\mathbb{R}^{n}$, $\Omega$ is invertible
on the pair of indices $(ij)$ and $(kl)$.

A subset of matrix space $\mathbb{R}^{n}\times\mathbb{R}^{n}$ which
becomes an interest in this article is the space of bivectors in $n$-dimensional
space, $\bigwedge^{2}\left(\mathbb{R}^{n}\right)$, namely a space
of antisymmetric matrix in $\mathbb{R}^{n}\times\mathbb{R}^{n}$.
It is spanned by orthonormal coordinate basis: $dx^{i}\wedge dx^{j}=dx^{i}\otimes dx^{j}-dx^{j}\otimes dx^{i},$
for $i,j=1,..,n$. Any element of $\bigwedge^{2}\left(\mathbb{R}^{n}\right)$
could be written as a linear combination of the basis: $\omega=\omega_{ij}dx^{i}\wedge dx^{j}.$
A bivector $\omega\in\bigwedge^{2}\left(\mathbb{R}^{n}\right)$ is
\textit{simple} if it can be written as a wedge product of two vectors
$u,v\in\mathbb{R}^{n}$:
\begin{equation}
\omega=u\wedge v.\label{eq:1}
\end{equation}
 Equivalently, if (\ref{eq:1}) is satisfied, then $\omega\wedge\omega=0$
\cite{Baez}. From the definition, it is clear that the coordinate
basis in bivector space is simple.

The following lemmas we proof in \cite{ICMNS} are important to derive
the main theorem in this article.\\

\textbf{Lemma 3.1 (Similarity of Simple Bivectors).}

\textit{Let $\omega\in\bigwedge^{2}\left(\mathbb{R}^{n}\right)$ be
a simple bivector such that:
\begin{equation}
\omega=u\wedge v=\frac{u\otimes v-\left(u\otimes v\right)^{T}}{2},\qquad u,v\in\mathbb{R}^{n}.\label{eq:req-1-1-1}
\end{equation}
The following transformation:}
\begin{equation}
\omega'=\Lambda\omega\Lambda^{T},\qquad\Lambda\in GL(n,\mathbb{R}),\label{eq:alhamdulillah}
\end{equation}
\textit{is the most general transformation preserving the simplicity
of a bivector, namely:
\begin{equation}
\omega'=u'\wedge v'.\label{eq:wei}
\end{equation}
All simple bivectors are similar up to transformation (\ref{eq:alhamdulillah}).}\\

The proof for Lemma 3.1 is sketched briefly as follows: The first
step is to obtain the most general transformation which preserve the
simplicity of matrix in $\mathbb{R}^{n}\times\mathbb{R}^{n}$ , namely
$\omega=u\otimes v$. The second step is to obtain the most general
transformations in which preserve (anti)-symmetricity of a matrix,
this transformation defines invariant subspaces in $\mathbb{R}^{n}\times\mathbb{R}^{n}$
, which are the space of symmetric and anti-symmetric matrices. Combining
these two results proves Lemma 3.1. The detailed derivation is carried
in \cite{ICMNS}.

The Frobenius inner product on $\mathbb{R}^{n}\times\mathbb{R}^{n}$
induced an equivalent inner product in the bivector space. This cause
the possibility to define matrix multiplication, and moreover, Lie
derivative in $\bigwedge^{2}\left(\mathbb{R}^{n}\right)$.\\

\textbf{Definition 3.1 (Coordinate Generators of (Special) Orthogonal
Group).}

\textit{The coordinate basis of bivector:
\[
\left\{ dx^{i}\wedge dx^{j}\right\} \in\bigwedge^{2}\mathbb{R}^{n}\qquad i,j=1,..n,
\]
is the }\textbf{\textit{coordinate generators}}\textit{ of special
orthogonal group $SO(n)$, or the coordinate basis of Lie algebra
$\mathfrak{so\left(n\right)},$ satisfying the following closed algebra
structure:
\begin{equation}
\left[dx^{i}\wedge dx^{j},dx^{j}\wedge dx^{k}\right]=\varepsilon^{ijk}\: dx^{i}\wedge dx^{k},\quad i,j=1,..,n.\label{eq:algebra}
\end{equation}
}\\
The coordinate generator $\left\{ dx^{i}\wedge dx^{j}\right\} $ is
simple and orthonormal\textit{, }but a generator of SO(n) in general
is not necessarily simple, as bivector in dimension $n\geq4$ in general
is not simple. Using Lemma 3.1, one could prove the following lemma.\\

\textbf{Lemma 3.2 (Similarity of Simple Generators of (Special) Orthogonal
Group).}

\textit{Let a set of non-coordinate basis of bivectors $\left\{ \hat{e}^{ij}\right\} $
for $i,j=1,..n$, satisfying the $\mathfrak{so(n)}$ algebra structure
relation (\ref{eq:algebra}), and:
\[
\hat{e}^{ij}=\hat{u}^{i}\wedge\hat{v}^{j},
\]
be a }\textbf{\textit{simple generators}}\textit{ of $SO(n)$, or
equivalently, simple basis of Lie algebra $\mathfrak{so\left(n\right)}$.
The following transformation: }
\begin{equation}
\hat{e}'^{ij}=\Lambda\hat{e}^{ij}\Lambda^{-1},\qquad\Lambda\in O(n),\label{eq:gauge}
\end{equation}
\textit{is the most general map preserving (simultaneously)  anti-symmetricity,
simplicity, and the closed algebra structure. All simple generators
of $\mathfrak{so\left(n\right)}$ are similar up to transformation
(\ref{eq:gauge}).}\\

The proof is rather direct. Using transformation (\ref{eq:alhamdulillah})
on (\ref{eq:algebra}) one could prove Lemma 3.2. The map (\ref{eq:gauge}),
is known as the adjoint action in on the algebra: 
\begin{eqnarray*}
\textrm{Ad}:\mathcal{G}\times\mathfrak{g} & \rightarrow & \mathcal{G}\\
\left(g,\omega\right) & \mapsto & \omega'=\textrm{Ad}_{g}\omega,
\end{eqnarray*}
which is also called as the similarity transformation.

As consequences of Lemma 3.2, we could conclude the following two
corollaries:\\

\textbf{Corollary 3.2A (Simple Representation of Lie Algebra $\mathfrak{so(3)}$
in $n$-Dimension).}

\textit{The subsets 
\begin{equation}
\left\{ dx^{a}\wedge dx^{b},dx^{b}\wedge dx^{c},dx^{c}\wedge dx^{a}\right\} \subset\left\{ dx^{i}\wedge dx^{j}\right\} ,\qquad i,j=1,..,n,\label{eq:yum}
\end{equation}
are coordinate generators of simple representation of $\mathfrak{so\left(3\right)}$
in $n$-dimension}.\textit{ Any sets $\left\{ \hat{e}^{ab},\hat{e}^{bc},\hat{e}^{ca}\right\} $
which are similar to (\ref{eq:yum}) up to a similarity transformation
(\ref{eq:gauge}) are non-coordinate generators of simple representation
of $\mathfrak{so\left(3\right)}$ in $n$-dimension}\textbf{\textit{. }}

\textit{Both (\ref{eq:yum}) and}\textbf{\textit{ }}\textit{$\left\{ \hat{e}^{ab},\hat{e}^{bc},\hat{e}^{ca}\right\} $
spans }$d\rho_{n}\left[\mathfrak{so\left(3\right)}\right]_{\textrm{sim}}$:\textit{
the space of simple representation of $\mathfrak{so(3)}$ in $n$-dimension.
All element of }$d\rho_{n}\left[\mathfrak{so\left(3\right)}\right]_{\textrm{sim}}$\textit{
are simple and similar up to a similarity transformation (\ref{eq:gauge}).
}\\

\textbf{Corollary 3.2B (Representation of Lie Algebra $\mathfrak{so(3)}$
in $n$-Dimension).}

\textit{Let a set of general bivectors (not necessarily simple) labeled
as follows: 
\[
\left\{ \hat{e}'^{ab},\hat{e}'^{bc},\hat{e}'^{ca}\right\} \in\bigwedge^{2}\mathbb{R}^{n},
\]
 satisfies the $\mathfrak{so\left(3\right)}$ algebra 
\begin{equation}
\left[\hat{e}'^{ab},\hat{e}'^{bc}\right]=\varepsilon^{abc}\hat{e}'^{ac},\label{eq:so3}
\end{equation}
 then the set defines a generator of representation of $\mathfrak{so\left(3\right)}$
in $n$-dimension. }

\textit{$\left\{ \hat{e}'^{ab},\hat{e}'^{bc},\hat{e}'^{ca}\right\} $
spans }$d\rho_{n}\left[\mathfrak{so\left(3\right)}\right]$:\textit{
the space of representation of $\mathfrak{so(3)}$ in $n$-dimension.
If there exist a group element $\Lambda\in O(n)$ such that $\left\{ \hat{e}'^{ab},\hat{e}'^{bc},\hat{e}'^{ca}\right\} $
is similar to $\left\{ \hat{e}^{ab},\hat{e}^{bc},\hat{e}^{ca}\right\} $,
then the generator $\left\{ \hat{e}'^{ab},\hat{e}'^{bc},\hat{e}'^{ca}\right\} $
is simple, and so do their linear combinations.}\\

As we have already mentioned in the previous section, the holonomy,
which is an element of a Lie group, could be obtained from relation
(\ref{eq:aa}), which is an exponential of the Lie algebra-valued
connection. But since holonomy, by its own right, is a group element
(which could be defined independently from the loop), it can be written
as an exponential of the Lie algebra element $J\in\mathfrak{so(n)}$
(not to be confused with the Lie algebra-valued connection $\boldsymbol{A}$)
:
\begin{equation}
U_{i}=\exp J_{i}=\exp\left|J_{i}\right|\hat{J}_{i}\in SO(n).\label{eq:taylor}
\end{equation}
One could conclude easily that the adjoint action (\ref{eq:gauge})
commutes with the exponential map.\textit{}\\

\textbf{Lemma 3.3 (Simple Representation of Lie Group $SO(3)$ in
$n$-Dimension).}

\textit{Let }$J\in d\rho_{n}\left[\mathfrak{so\left(3\right)}\right]$\textit{
be an element of representation of $\mathfrak{so\left(3\right)}$
in $n$-dimension. The element of representation of the group SO(3)
in n-dimension, }$\rho_{n}(SO\left(3\right))$,\textit{ can be obtained
from the exponential map:
\begin{equation}
U=\exp J\in\rho_{n}(SO\left(3\right)).\label{eq:d}
\end{equation}
 If $J$ is simple, then any element of simple representation of $SO\left(3\right)$
in n-dimension, $\rho_{n}\left(SO\left(3\right)\right)_{\textrm{sim}}$,
can always be written as follows:}
\begin{equation}
U=I_{n\times n}+\hat{J}\sin\theta+\hat{J}^{2}\left(1-\cos\theta\right)\in\rho_{n}\left(SO\left(3\right)\right)_{\textrm{sim}},\qquad\theta=\left|J\right|\label{eq:waw}
\end{equation}
\textit{up to the similarity transformation (\ref{eq:gauge}).}\\

The proof is the following. Using Corollary 3.2A, all simple bivector
in $n$-dimension are similar up to a similarity transformation (\ref{eq:gauge}),
namely, there always exist $\Lambda\in SO(n)$ which bring $J$ to
$J'\in\mathfrak{so(n).}$ Since the similarity transformation commutes
with the exponential map, there always exists\textit{ }$\Lambda\in SO(n)$
which bring (\ref{eq:d}) to the form of (\ref{eq:waw}), which proof
the previous lemma. Lemma 3.3 is particularly important in deriving
our main theorem of this article:\\

\textbf{Theorem 3.2 (Trace Relation of Gauge Group $SO(n)$: The $SO(3)$
Case)}

\textit{Let $U_{1}$, $U_{2}$, $U_{3}$ $\in$ $SO\left(n\right)$.
If $U_{1}$, $U_{2}$, $U_{3}$ satisfy the following condition:}
\begin{enumerate}
\item \textit{$U_{1}$, $U_{2}$, $U_{3}$ $\in$ $\rho_{n}\left[SO\left(3\right)\right]$
$\subset$ $SO\left(n\right)$, that is, they are elements of common
SO(3) subgroup representation in $n$-dimension as in }\textbf{\textit{Lemma
3.3}}\textit{,}
\item \textit{$U_{1}$, $U_{2}$, $U_{3}$ are simple, that is, they are
exponential of simple Lie Algebra, and}
\item \textit{$U_{1}$, $U_{2}$, $U_{3}$ satisfy the Bianchi identity
(\ref{eq:bifinite}), }
\end{enumerate}
\textit{then the trace relation or the contracted Bianchi Identity
in the form of (\ref{eq:3.1}) gives the angle relation (\ref{eq:3.2}).}

\textbf{Proof}. From requirements (1) and (2), together with Corollary
3.2B, one can conclude that \textit{$U_{1}$, $U_{2}$, $U_{3}$}
$\in$\textit{ }$\rho_{n}\left(SO\left(3\right)\right)_{\textrm{sim}}$.
From Lemma 3.3, every element of $\rho_{n}\left(SO\left(3\right)\right)_{\textrm{sim}}$
could always be written as (\ref{eq:waw}). Taking the trace of the
second Bianchi Identity as in (\ref{eq:3.1}) and using the half angle
formula gives the angle relation (\ref{eq:3.2}).\\

It must be kept in mind that in general, the trace relation does not
gives the angle relation, only the one which satisfies point 1 and
2 of Theorem 3.2 does.

\subsection{The case of SU(2) as the gauge group}

The relation (\ref{eq:3.2}) is somewhat unsatisfying; in addition
that it contains\textit{ two} solutions, the dihedral angle $\theta$'s
are \textit{half} of the angle of rotation, while the planar angle
$\phi$ is fixed. As a comparison, the following is the original dihedral
angle relation: \textit{
\begin{equation}
\cos\theta_{3}=\cos\theta_{1}\cos\theta_{2}-\cos\phi_{12}\sin\theta_{1}\sin\theta_{2},\label{eq:realdihed}
\end{equation}
}with the geometrical interpretation in 3-dimension given in FIG.
2. 
\begin{figure}
\begin{centering}
\includegraphics[scale=0.6]{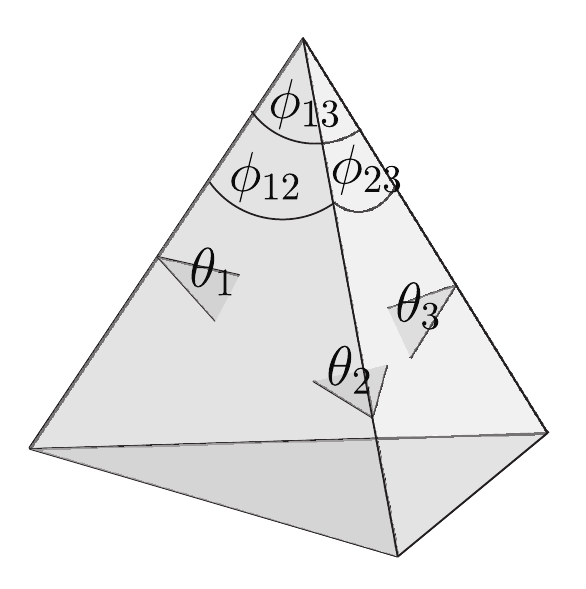}
\par\end{centering}

\caption{Types of angles inside a tetrahedron: the 2-dimensional angle $\phi_{ij}$
are the angle between segments, located on a point where these segments
meet; and the 3-dimensional 'dihedral' angle $\theta_{i}$ are the
angle between triangles, located on segments. These angles satisfy
'dihedral' angle relation in (2+1)-dimension.}
\end{figure}
Interestingly, if we choose the gauge group on Theorem 3.1 to be SU(2)
instead of SO(3) so that $U_{i}\in SU(2)$, the trace relation gives
exactly (\ref{eq:realdihed}).\\

\textbf{Theorem 3.3 (Trace Relation of Gauge Group SU(2)).}

\textit{Given elements of group $U_{i}\in SU(2)$, the trace relation
or the contracted Bianchi Identity in the form of (\ref{eq:3.1})
gives the dihedral angle relation (\ref{eq:realdihed}), with $\theta_{i}$
are the rotation angles of $U_{i}$} \textit{and $\phi_{12}$ is the
angle between plane of rotation of $U_{1}$ and $U_{2}$.}

\textbf{Proof.} The proof of this is direct and straightforward, realizing
that any $U_{i}\in SU(2)$ can always be written as:\textit{
\begin{equation}
U_{i}=I_{2\times2}\cos\theta_{i}+\hat{J}_{i}\sin\theta_{i},\qquad\hat{J}_{i}\in\mathfrak{su(2)},\quad\theta_{i}=\left|J_{i}\right|.\label{eq:3.3-1}
\end{equation}
}Using the following relations for $J_{i}=\vec{J}_{i}\cdot\hat{\sigma}\in\mathfrak{su(2)},$
with $\vec{J}_{i}\in\mathbb{R}^{3}$ and $\hat{\sigma}$ are generator
of $\mathfrak{su(2)}$: $\textrm{tr}\left(J_{i}J_{j}\right)=-2\left\langle \vec{J}_{i},\vec{J}_{j}\right\rangle $,
one could obtain dihedral relation (\ref{eq:realdihed}).\\

In fact, one could obtain (\ref{eq:realdihed}) with a slight modification
of the algebra of $\mathfrak{so(3)}$ as follows: Let $J_{i}=\vec{J}_{i}\cdot\hat{l}\in\mathfrak{so(3)},$
with $\hat{l}$ is the generator of $\mathfrak{so(3)}$. Let us define
the transformation on the generator as follows: $l_{a}\mapsto\tau_{a}=2l_{a},$
where the algebra of $\left\{ \hat{\tau}\right\} $ is: 
\begin{equation}
\left[\tau_{a},\tau_{b}\right]=-2\varepsilon_{abc}\tau_{c},\label{eq:su2}
\end{equation}
instead of (\ref{eq:so3}). Using the new generator, an element of
$\mathfrak{so(3)}$ could be written as $T=\vec{T}_{i}\cdot\tau_{a}\in\mathfrak{so(3)}$,
and moreover its corresponding group element of $SO(3)$ could be
written as:
\begin{equation}
U_{i}=I_{3\times3}+\hat{J}_{i}\sin\,2\theta_{i}+\hat{J}_{i}^{2}\left(1-\cos\,2\theta_{i}\right),\qquad J_{i}\in\mathfrak{so(3)}.\label{eq:rite}
\end{equation}
(\ref{eq:su2}) is exactly the algebra relation of $\mathfrak{su(2)}$.
Specifically, $\tau_{a}$ are the irreducible representation of generator
of $SU(2)$ in 3-dimension:
\begin{equation}
d\rho_{3}:\mathfrak{su(2)}\rightarrow\mathfrak{so(3)}\subset\mathfrak{gl(n,R),}\label{eq:real}
\end{equation}
since there exist a similarity transformation $d\rho_{3}\left(\hat{\sigma}\right)=\Lambda\hat{\tau}\Lambda^{-1}$
by $\Lambda\in SU(2)$ which bring $\tau_{a}$ to the usual basis
of irreducible representation of generator of $SU(2)$ in 3-dimension,
namely, $d\rho_{3}\left(\hat{\sigma}\right)$. Using (\ref{eq:rite}),
the trace relation gives:
\begin{equation}
\cos\theta_{3}=\pm\left(\cos\theta_{1}\cos\theta_{2}-\cos\phi_{12}\sin\theta_{1}\sin\theta_{2}\right),\label{eq:huff-1}
\end{equation}
where one of the solution is exactly the dihedral angle relation (\ref{eq:realdihed}).

As a generalization to Theorem 3.3, we rederive Theorem 3.2, but now,
the condition for gauge group $\rho_{n}\left[SO(3)\right]$ is replaced
by $\rho_{n}\left[SU\left(2\right)\right]$:\\

\textbf{Theorem 3.4 (Trace Relation of Gauge Group SO(n): The SU(2)
Case)}

\textit{Let $U_{1}$, $U_{2}$, $U_{3}$ $\in$ $SO\left(n\right)$.
If $U_{1}$, $U_{2}$, $U_{3}$ satisfy the following condition:}
\begin{enumerate}
\item \textit{$U_{1}$, $U_{2}$, $U_{3}$ $\in$ $\rho_{n}\left[SU\left(2\right)\right]$
$\subset$ $SO\left(n\right)$, that is, they are elements of common
SU(2) subgroup representation in $n$-dimension as in }\textbf{\textit{Lemma
3.3}}\textit{,}
\item \textit{$U_{1}$, $U_{2}$, $U_{3}$ are simple, and}
\item \textit{$U_{1}$, $U_{2}$, $U_{3}$ satisfy the Bianchi identity
(\ref{eq:bifinite}), }
\end{enumerate}
\textit{then the trace relation or the contracted Bianchi Identity
in the form of (\ref{eq:3.1}) gives the angle relation (\ref{eq:huff-1}).}

\textbf{Proof}. The proof is started from writing the $\mathfrak{su(2)}$
real representation in 3-dimension as in (\ref{eq:real}). Since (\ref{eq:real})
sends $T\in\mathfrak{su(2)}$ to $\mathfrak{so(3)},$ the next step
for proving Theorem 3.4 can be carried exactly as the proof for Theorem
3.2.\\

It is quite intriguing the fact that the one which gives the dihedral
angle relation, which is an aspect of Euclidean 3-dimensional space,
is the SU(2) group instead of SO(3). It is well-known that SU(2) is
the Spin(3) group which double covers the group SO(3). As a manifold,
SO(3), topologically, is isomorphic to $\mathbb{RP}^{3}$: the real
projective space in 3-dimension, which is not simply-connected. In
the other hand, SU(2) is a 3-sphere, which is topologically simpler
than $\mathbb{RP}^{3}$. The fact that the dihedral angle relation
(\ref{eq:huff-1}) is exactly the spherical law of cosine if one takes
$\bar{\phi}=2\pi-\phi$ may provide the reason why it is more natural
to use SU(2) instead of SO(3) as the gauge-group for spatial geometries,
in particular, Regge geometries. We will discuss this in the last
section. Given the advantage of SU(2) as a gauge group, it is relevant
to continue to Section C where the geometrical meaning of the spherical
law of cosine is discussed.

\subsection{Geometrical Interpretation: Spherical Law of Cosine and Dihedral
Angle Relation}

Let us return to relation (\ref{eq:taylor}) in Section 2A. Rewriting
$\left|J_{i}\right|$ as $\theta_{i}$, and approximating the dihedral
angle relation (\ref{eq:realdihed}) for small angle of rotation,
one obtains: 
\begin{equation}
\theta_{3}^{2}\cong\theta_{1}^{2}+\theta_{2}^{2}+2\cos\phi_{12}\theta_{1}\theta_{2},\label{eq:1-1}
\end{equation}
or:
\begin{equation}
\left|J_{3}\right|^{2}=\left|J_{1}\right|^{2}+\left|J_{2}\right|^{2}+2\left|J_{1}\right|^{2}\left|J_{2}\right|^{2}\cos\phi_{12}.\label{eq:trianglein}
\end{equation}
(\ref{eq:trianglein}) could arise from the following inner product:
\begin{equation}
\left\langle J_{3},J_{3}\right\rangle =\left\langle J_{1}+J_{2},J_{1}+J_{2}\right\rangle \label{eq:hayah}
\end{equation}
of the closure condition $J_{1}+J_{2}+J_{3}=0,$ which has the same
form with Bianchi Identity (\ref{eq:flat2}) in the nearly-flat case
with finite loop and infinitesimal curvature, given $\intop_{S_{i}}\boldsymbol{F}=J_{i}$.

Studying the discrete nearly-flat case, one could have the intuition
that the trace of a matrix can be interpreted as a generalization
of an 'inner product'. In fact, this had been introduced not so recently
by von Neumann, in the statistical formulation of quantum mechanics
\cite{Neumann}, where a pure state $\left|\psi\right\rangle $ is
written by a density matrix $\left|\psi\right\rangle \left\langle \psi\right|,$
a simple and symmetric (possibly, infinite-dimensional) matrix. Its
inner product in the complex Hilbert space is exactly:
\[
\left.\left\langle \psi\right|\psi\right\rangle =\textrm{tr}\left|\psi\right\rangle \left\langle \psi\right|.
\]
A statistical generalization of a pure state is a mixed state:
\[
\rho_{\psi}=\sum_{i}\lambda_{i}\left|\psi_{i}\right\rangle \left\langle \psi_{i}\right|=\int dx\,\lambda\left(x\right)\left|\psi\left(x\right)\right\rangle \left\langle \psi\left(x\right)\right|,
\]
a linear combination of simple matrices, which in general is not simple.
The (semi)-positive definiteness of the trace, and hence the possibility
to use it as an inner product in complex Hilbert space, is a consequence
of the symmetricity of the density matrices $\rho_{\psi}$. In general,
trace of an arbitrary matrix is not positive definite. We adopt the
generalization of 'inner product' for the trace of SO(n) group, where
the elements are not symmetric. The corresponding Lie algebra \textit{$\mathfrak{so\left(n\right)}$},
which are the infinitesimal rotations described by antisymmetric matrices,
gives zero trace, thus the usual inner product is used to obtain the
usual norm.

The condition that the elements of SO(n) needs to be simple representation
of a common SU(2) in $n$-dimension can be thought of as the existence
of an embedding of a 3-sphere on an $n$-dimensional manifold SO(n).
\[
\rho_{n}:SU\left(2\right)\rightarrow\rho_{n}\left(SU\left(2\right)\right)_{\textrm{sim}}\subset SO(n).
\]
In the next paragraph, we will show that the trace relation of SO(n)
satisfying Theorem 3.4, is a condition for the existence of a spherical
triangle on the great 2-sphere of SU(2).

With the 'inner product' interpretation of the trace relation, the
geometrical interpretation of the condition described by Theorem 3.4
is given as follows. The group $U_{\gamma_{i}}$ on Theorem 3.4 could
be written in the plane-angle representation: 
\[
U_{i}=\exp\hat{J}_{i}\theta_{i},\qquad\left|\hat{J}_{i}\right|=1,\quad\theta_{i}\in\mathbb{R}^{+}.
\]
Thus there exists a spherical triangle with length $\theta_{1},\theta_{2},\theta_{3}$
in a sphere $\mathbb{S}^{2}\subset$ $\bigwedge^{2}\mathbb{R}^{n}\sim\mathfrak{g}$,
where the center of the sphere is located in the origin $\mathcal{O}$
of $\bigwedge^{2}\mathbb{R}^{n}$. See FIG. 3a. 
\begin{figure}
\begin{centering}
\includegraphics[scale=0.6]{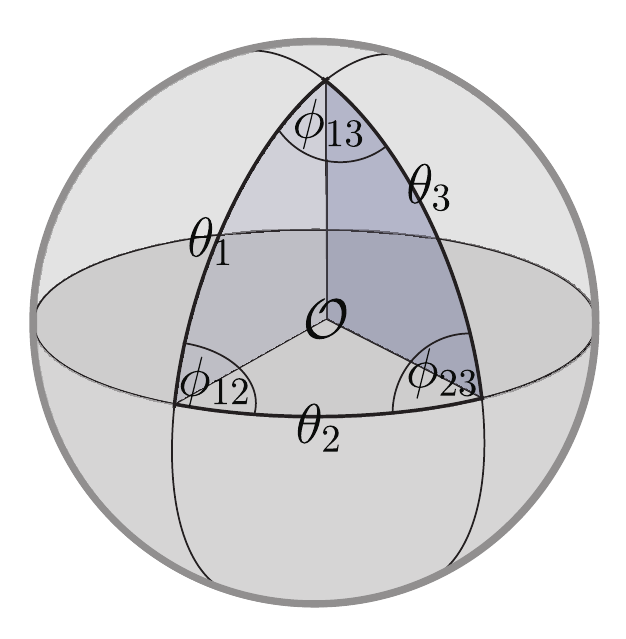}
\par\end{centering}

\caption{A spherical triangle on $\mathbb{S}^{2}$, constructed by the intersection
of three great circles of the 2-sphere. The sides $\theta_{i}$ are
the arc of the great circles, and the angles $\phi_{ij}$ is defined
by the intersection between two of the arcs. The sum of the $\phi_{ij}$
inside the triangle is larger than $\pi$. Viewing the $\mathbb{S}^{2}$
as embedded in $\mathbb{R}^{3}$, point $\mathcal{O}$ inside the
3-ball enclosed by the 2-sphere, together with the three points of
the spherical triangles describe a trihedron $J_{i}$, $i=1,2,3$,
which clearly satisfy the 'dihedral' angle relation. For small $\theta_{i}$,
the spherical triangle is approximated by a  flat triangle. }

\end{figure}

For a case where the angle $\theta_{1},\theta_{2},\theta_{3}$ is
small, relation (\ref{eq:1-1}) becomes the closure condition (\ref{eq:flat2}),
$J_{1}+J_{2}+J_{3}\approx0,$ which is the Bianchi Identity in the
nearly-flat case with finite loop and infinitesimal curvature. It
could be written as:
\[
\hat{J}_{1}\theta_{1}+\hat{J}_{2}\theta_{2}+\hat{J}_{3}\theta_{3}=0,\qquad\left|\hat{J}_{i}\right|=1,\quad\theta_{i}\in\mathbb{R}^{+}.
\]
Then, using the theorem discovered by Minkowski \cite{Minkowski},
there exist a triangle with length $\theta_{1},\theta_{2},\theta_{3}$
in $\bigwedge^{2}\mathbb{R}^{n}\sim\mathfrak{g}$, where the center
is located in the origin $\mathcal{O}$ of $\bigwedge^{2}\mathbb{R}^{n}$.
Thus the spherical triangle is approximated by a flat triangle. 

As a conclusion for this section, we have already obtained Theorem
3.2 and particularly, Theorem 3.4. This theorem is important to defend
the argument that the holonomy group of an oriented, $n$-dimensional
simplicial complex in Euclidean Regge Calculus needs to be SO(n) plus
some certain condition, which will be clear in the next section.

\section{4D case}

For Euclidean gravity, one is interested, particularly, in the map:
\begin{equation}
\rho_{4}:SU\left(2\right)\rightarrow\rho_{4}\left(SU\left(2\right)\right)_{\textrm{sim}}\subset SO(4).\label{eq:rep}
\end{equation}
The group SO(4) is topologically isomorphic to $\mathbb{RP}^{3}\times\mathbb{S}^{3}$,
thus the representation map (\ref{eq:rep}) is an immersion of a 3-sphere
into $\mathbb{RP}^{3}\times\mathbb{S}^{3}$. 

In 3-dimension and lower, all bivectors are simple, but this is not
the case in four and larger dimension. Particularly in $\bigwedge^{2}\left(\mathbb{R}^{4}\right)\sim\mathfrak{so(4)}$,
one could define the notion of (anti) self-duality, since the Hodge
star operator sends 2-forms to 2-forms. A bivector $\omega_{\pm}\in\mathfrak{so(4)}$
is a (anti) self-dual bivector if it satisfies $\omega_{\pm}=\pm\star\omega_{\pm}.$
Any bivector $\omega\in\mathfrak{so(4)}$ can be decomposed into its
self-dual and anti self-dual parts, such that $\omega=\omega_{+}+\omega_{-}.$
Using the notion of self-duality, it is possible to write the components
of any matrix $^{4}U\in SO(4)$ in terms of trigonometric functions
as follows.\\

\textbf{Theorem 4.1 (Elements of SO(4))}

\textit{Let $J$ be an element of }$\mathfrak{so(4)}$\textit{, written
in its self and anti self-dual parts as follows:
\begin{equation}
J=J_{+}+J_{-},\quad\left(J_{\pm}\right)_{\mu\nu}\left\{ \begin{array}{ccc}
\left(J_{\pm}\right)_{00} & = & 0\\
\left(J_{\pm}\right)_{0i} & = & \pm\left(j_{\pm}\right)_{i}\\
\left(J_{\pm}\right)_{i0} & = & \mp\left(j_{\pm}\right)_{i}\\
\left(J_{\pm}\right)_{ij} & = & \varepsilon_{ijk}\left(j_{\pm}\right)^{k}
\end{array}\right.,\qquad\left(j_{\pm}\right)_{i}\in\mathbb{R}^{+}\label{eq:hue}
\end{equation}
An element of SO(4) can be written as:}

\textit{
\begin{eqnarray}
^{4}U & = & U_{+}U_{-},\label{eq:so4}
\end{eqnarray}
with:
\[
U_{\pm}=\exp\left(J_{\pm}\right)=\exp\left(\hat{J}_{\pm}\left|J_{\pm}\right|\right)\in\rho_{4}\left[SO\left(3\right)^{\pm}\right],
\]
are the self and anti self-dual part of $^{4}U$. Moreover, it could
be written as:
\begin{eqnarray}
^{4}U & = & I_{4\times4}\cos\varphi^{+}\cos\varphi^{-}+\hat{J}^{-}\cos\varphi^{+}\sin\varphi^{-}+\hat{J}^{+}\sin\varphi^{+}\cos\varphi^{-}+\hat{J}^{+}\hat{J}^{-}\sin\varphi^{+}\sin\varphi^{-},\label{eq:huff}
\end{eqnarray}
with $\varphi^{\pm}=\left|J_{\pm}\right|=\left|j_{\pm}\right|$.}

\textbf{Proof.} The proof for Theorem 4.1 is straightforward: Let
$J\in\mathfrak{so(4)}$ be decomposed into its (anti) self-dual parts
$J^{\pm}.$ Since the self-dual and anti self-dual parts commute,
the element of SO(4) can always be written as (\ref{eq:so4}) using
the exponential map. By Taylor expansion, (\ref{eq:so4}) can be written
as (\ref{eq:huff}).\\

Now we are ready to classify all types of rotations in 4-dimensional
Euclidean space. The classification is based on the corresponding
Lie algebra of $^{4}U$, which can be obtained from the inverse of
relation (\ref{eq:taylor}), namely $J=\ln U.$

It is a well-known fact that any $n\times n$ anti-symmetric matrix
is similar with an anti-symmetric, block diagonal matrix by an orthogonal
transformation O(n). Using this result, a $4\times4$ anti-symmetric
matrix $J\in\mathfrak{so(4)}$, can be written as an anti-symmetric
$2\times2$ block diagonal matrix $J'$ as follows:
\begin{equation}
\Lambda J\Lambda^{-1}=J',\qquad J'=\left[\begin{array}{cc}
\lambda_{+}\sigma_{z} & \mathbf{0}_{2\times2}\\
\mathbf{0}_{2\times2} & \lambda_{-}\sigma_{z}
\end{array}\right],\;\lambda_{\pm}\in\mathbb{R},\label{eq:dem}
\end{equation}
 by an orthogonal similarity transformation $\Lambda\in O(4).$ $\sigma_{z}$
is the $z$-components of Pauli matrix. Both $\lambda_{+}\sigma_{z}$
and $\lambda_{-}\sigma_{z}$ describe geometrically the invariant
planes of $^{4}U$, where $\lambda_{+}\sigma_{z}$ is fixed by a rotation
of $\lambda_{-}\sigma_{z}$ and vice-versa. One could arrive to the
conclusion that a bivector in 4-dimension, or an element of $\mathfrak{so(4)},$
can be written as a direct sum of two simple bivectors $\lambda_{+}\sigma_{z}$
and $\lambda_{-}\sigma_{z}$, up to an orthogonal transformation.
Using these fixed planes, one can defined the following classification
for the rotations generated by \textit{$J\in\mathfrak{so(4)}$}:
\begin{enumerate}
\item $^{4}U$ describe a \textbf{simple (or single) rotation}, if one of
the plane have zero norm: either $\lambda_{+}=0,$ or $\lambda_{-}=0$. 
\item $^{4}U$ describe a \textbf{double (or Clifford) rotation}, if $\lambda_{+}\neq\lambda_{-}.$
\item $^{4}U$ describe an\textbf{ isoclinic rotation}, if $\lambda_{+}=\lambda_{-}.$ 
\end{enumerate}
Let us reviewed each case and see if it is possible to relate the
classification with the self-duality of $J\in\mathfrak{so(4)}$. One
could recall a remarkable relation between self-duality and simplicity
of a bivector in 4-dimension\textit{: a bivector $J\in\mathfrak{so(4)}$
is simple if and only if the norm of the self-dual and anti self-dual
parts are equal}. A direct calculation on the diagonalization of a
$4\times4$ anti-symmetric matrix in the form of (\ref{eq:hue}),
gives the eigenvalues $\lambda_{\pm}$ of (\ref{eq:dem}):
\begin{equation}
\lambda_{\pm}=\left|j_{+}\right|\pm\left|j_{-}\right|.\label{eq:eigen}
\end{equation}
Using this fact, we could classify the elements of $J\in\mathfrak{so(4)}$
as follows.
\begin{enumerate}
\item \textbf{The case $j_{+}=\pm j_{-}$}. \\
This automatically gives the simplicity condition $\left|j_{+}\right|=\left|j_{-}\right|.$
Therefore, $J$ is simple and not self-dual. The constraint \textbf{$\left(j_{+}\right)_{i}=\pm\left(j_{-}\right)_{i}$}
defines a three-dimensional subspace $\Omega\subset\mathfrak{so(4)}$.
The subspace is spanned by generators $\left\{ l\right\} $ satisfying
the $\mathfrak{so(3)}$ (or $\mathfrak{su(2)}$) algebra relation.
It is clear that $\Omega$ is isomorphic to $d\rho_{n}\left(\mathfrak{so(3)}\right)_{\textrm{sim}}$,
that is, the space of simple (irreducible) representation of $\mathfrak{so(3)}$
in 4-dimension. There exist only a single non-zero plane of rotation,
which is either $\lambda_{+}\sigma_{z}=2\left|j\right|\sigma_{z}$
or $\lambda_{-}\sigma_{z}=2\left|j\right|\sigma_{z}$, depending on
the $\pm$ sign. The exponential map of such elements, say $U\in\rho_{n}\left(SO(3)\right)_{\textrm{sim}}$
can be obtained from Lemma 3.3. Any element $U\in\rho_{n}\left(SO(3)\right)_{\textrm{sim}}$
describe the \textit{simple} rotation in 4-dimension.
\item \textbf{The case $j_{+}=0,$ or $j_{-}=0$.}\\
Since the norms are not equal, $J$ is not simple, but is self or
anti self-dual. The constraint defines a three-dimensional subspace
$\Sigma_{\pm}\subset\mathfrak{so(4)}$. Nevertheless, the degrees
of freedom is three, spanned by generators $\left\{ J^{\pm}\right\} $
satisfying the $\mathfrak{so(3)}$ (or $\mathfrak{su(2)}$) algebra
relation. This is the space of semi-simple representation of $\mathfrak{so(3)}$
in 4-dimension. The planes of rotation have equal norm, which are
either $\lambda_{\pm}\sigma_{z}=\left|j_{+}\right|\sigma_{z}$ or
$\lambda_{\pm}\sigma_{z}=\left|j_{-}\right|\sigma_{z}$ , depending
on which part is zero. Thus the exponential map of element of $\Sigma_{\pm}$
describe the left or right \textit{isoclinic} rotation. 
\item \textbf{The case $\left|j_{+}\right|=\left|j_{-}\right|$.}\\
This caused $J$ to be simple, and the constraint $\left|j_{+}\right|=\left|j_{-}\right|$
defines a five-dimensional subspace $\mathfrak{so(4)}_{\textrm{sim}}\subset\mathfrak{so(4)}$,
which is the space of simple bivectors in 4-dimension. In general,
elements of $\mathfrak{so(4)}_{\textrm{sim}}$ is \textit{not} a representation
of $\mathfrak{so(3)}$ in 4-dimension, such that $\Omega\subset\mathfrak{so(4)}_{\textrm{sim}}$.
Nevertheless, they have similar single plane of rotation as in the
first case and therefore describe \textit{simple} rotations.
\item \textbf{The case $\left|j_{+}\right|\neq\left|j_{-}\right|.$} \\
This is the most general case where $J$ is semi-simple. The exponential
map of such elements has two distinct planes $\lambda_{\pm}\sigma_{z}$
satisfying (\ref{eq:eigen}), describing \textit{double} (or Clifford)
rotation in 4-dimension.
\end{enumerate}
According to the (anti) self-dual pair $J^{\pm}$, we could conclude
the following fact for rotations in 4-dimensional Euclidean space:\\

\textbf{Corollary 4.1 (Classification of rotations in 4-dimensional
Euclidean Space)}

\textit{Given $U\in SO(4),$ satisfying $U=\exp J$ with $J\in\mathfrak{so(4)},$
the following statements are true:}
\begin{enumerate}
\item \textit{If $J$ is a simple bivector, then $U$ is a simple (or single)
rotation.}
\item \textit{If $J$ is either self-dual or anti self-dual, then $U$ is
either  a left or right isoclinic rotation.}
\item \textit{If $J$ is semi-simple, then $U$ is a double (or Clifford)
rotation.}\\

\end{enumerate}
Given three elements of group $U_{1}$, $U_{2}$, $U_{3}$ $\in\Omega$
of Case 1, they will automatically satisfy point 1 and 2 from Theorem
3.2 and 3.4. Moreover, if they satisfy the Bianchi Identity (\ref{eq:bifinite}),
their trace relation will gives angle relations. In the last chapter,
we will show that the condition $\left|J^{+}\right|=\left|J^{-}\right|,$
is crucial to obtain a (simplicial) complex in Regge gravity.

\section{Comments on Regge Geometries}

At the end of this article, we apply the theorems concerning the elements
of group to holonomies, that is, the group variables attached on the
loops. The theorem gives condition on the loops, as well as on the
holonomies, with the geometrical interpretations will also follows.

\subsection{From Second Order to First Order Formulation}

The spacetime in Regge gravity is modeled as a $4$-dimensional manifold
discretized by $4$-dimensional simplicial complex, nevertheless the
construction is valid for any dimension $n$. Each simplex in the
complex is uniquely defined by its edges length $\left|l_{m}\right|\in\mathbb{R}$.
Given these edges length, all higher geometrical variables such as
the volume-form of the sub-simplex, as well as the angles between
each two of them, are known. The curvature of the discretized manifold,
defined as the deficit angles $\delta\theta=2\pi-\sum_{m}\theta_{m},$
are concentrated on the $\left(n-2\right)$-simplices, called as hinges. 

Following the procedure described in \cite{Sato}, the Riemann tensor
for each hinge $h_{i}$ can be written as:
\begin{equation}
\boldsymbol{R}_{h_{i}}=\delta\theta_{i}\hat{\omega}_{h_{i}}\otimes\hat{\omega}_{h_{i}},\label{eq:5.1}
\end{equation}
with $\delta\theta_{i}$ is deficit angle on hinge $h_{i}$ and $\hat{\omega}_{h_{i}}$
is a generator of rotation, defined as a unit bivector, Hodge-dual
to the direction of hinge $h_{i}$ as follows:
\begin{equation}
\hat{\omega}_{h_{i}}=\frac{\star\left(l_{i_{1}}\wedge..\wedge l_{i_{n-2}}\right)}{\left|l_{i_{1}}\wedge..\wedge l_{i_{n-2}}\right|}.\label{eq:hi}
\end{equation}
Using the fact that the deficit angle is proportional to the product
of the modulus of rotation and area enclosed by the loop, $\delta\theta_{i}\sim\left|\omega_{h_{i}}\right|\left|\alpha_{h_{i}}\right|$,
(\ref{eq:5.1}) can be written as:
\begin{equation}
\boldsymbol{R}_{h_{i}}=\kappa\,\omega_{h_{i}}\otimes\alpha_{h_{i}},\label{eq:5.2}
\end{equation}
with $\alpha_{h_{i}}$ is the (infinitesimal) loop orientation which
is effectively a plane, $\omega_{h_{i}}$ is the infinitesimal rotation
parallel to $\alpha_{h_{i}}$, and $\kappa$ is a constant \cite{Miller}.
(\ref{eq:hi}) guarantees the simplicity of $\omega_{h_{i}}$ and
$\alpha_{h_{i}}$. Therefore, the Riemann tensor of a discretized
Regge manifold where the curvature is concentrated on the hinge is:
\[
\boldsymbol{R}\left(\boldsymbol{x}\right)=\rho\left(\boldsymbol{x},\boldsymbol{x}_{h_{i}}\right)\boldsymbol{R}_{h_{i}},
\]
with $\rho\left(\boldsymbol{x},\boldsymbol{x}_{h_{i}}\right)$ is
the support for $\boldsymbol{R}_{h_{i}}$ which are constant along
the hinges but vanish elsewhere.

The full contraction of Riemann tensor (\ref{eq:5.1}) is the Ricci
scalar for each hinge, written as:
\begin{equation}
R_{h_{i}}=\kappa\,\textrm{tr}\left(\omega_{h_{i}}^{T}\alpha_{h_{i}}\right)=2\delta\theta_{i}.\label{eq:ricci}
\end{equation}
Moreover, the action of general relativity is $S=\int\star R\left(\boldsymbol{x}\right)=\int R\left(\boldsymbol{x}\right)\textrm{vol}$,
such that inserting (\ref{eq:ricci}) gives:
\begin{equation}
S=\intop\rho\left(\boldsymbol{x},\boldsymbol{x}_{h_{i}}\right)R_{h_{i}}\textrm{vol}=\sum_{i}R_{h_{i}}\underset{\left|V_{h_{i}}\right|}{\underbrace{\intop\rho\left(\boldsymbol{x},\boldsymbol{x}_{h_{i}}\right)\textrm{vol}}}=2\sum_{i}\delta\theta_{i}\left|V_{h_{i}}\right|,\label{eq:reggeac}
\end{equation}
with $\left|V_{h_{i}}\right|$ is the measure of hinge $h_{i}$. (\ref{eq:reggeac})
is exactly the Regge action \cite{Regge1}. 

The condition (\ref{eq:hi}) guarantees the simplicity of $\alpha_{h_{i}}$
and $\omega_{h_{i}}$, which in turns guarantees that the Riemann
tensor (\ref{eq:5.2}) arise from a simplicial complex. In general,
a non-simple bivectors do not have a well-defined and concrete geometrical
interpretation, for example, it will be impossible to define a vector
parallel (and perpendicular) to a non-simple bivector. 

Let us proceed to the first order formulation, where the Riemann and
the curvature 2-form is related by a local trivialization $\boldsymbol{R}=e\left(\boldsymbol{F}\right)$
or:
\begin{equation}
R_{aijb}=e_{a}^{\;\, I}e_{b}^{\;\, J}F_{IijJ},\label{eq:1st}
\end{equation}
where $e_{a}^{\:\, I}$ is orthogonal. (\ref{eq:1st}) could be written
simply as $\boldsymbol{R}=e\boldsymbol{F}e^{T}=e\boldsymbol{F}e^{-1}$
by the orthogonality of $e$. But from Lemma 3.2, (\ref{eq:1st})
preserve the simplicity of a bivector, and this caused the curvature
2-form on hinge $h_{i}$ to satisfy:
\begin{equation}
\boldsymbol{F}_{h_{i}}=\kappa\,\overset{\omega_{h_{i}}'}{\overbrace{e^{-1}\omega_{h_{i}}e}}\otimes\alpha_{h_{i}}=\underset{\delta\theta_{i}}{\underbrace{\kappa\left|\alpha_{h_{i}}\right|\left|\omega_{h_{i}}\right|}}\hat{\omega}_{h_{i}}^{'}\otimes\hat{\omega}_{h_{i}},\label{eq:fregge}
\end{equation}
where $\omega_{h_{i}}'$ is also simple. One could notice that $\alpha_{h_{i}}$
and $\omega_{h_{i}}'$ do not necessarily need to be parallel to each
other. The curvature 2-form of discrete gauge gravity is: 
\[
\boldsymbol{F}\left(\boldsymbol{x}\right)=\rho\left(\boldsymbol{x},\boldsymbol{x}_{h_{i}}\right)\boldsymbol{F}_{h_{i}},
\]
and the first order Regge action is unchanged since $e$ is orthogonal.

For the reason concerning the fundamental discreteness in spacetime
explained in the end of Chapter 2, one needs to apply a regularization
procedure to Regge gravity, particularly, on the connection and curvature
2-form. The curvature 2-form is regularized into its corresponding
holonomy on the hinge by relation (\ref{eq:aa}). If the loop $\gamma_{i}=\partial S_{i}$
only circles a single hinge $h_{i}$, then:
\begin{equation}
U_{S_{i}}=\hat{P}\exp\underset{J_{i}}{\underbrace{\kappa\left|\alpha_{h_{i}}\right|\omega_{h_{i}}'}}\underset{1}{\underbrace{\int_{S_{i}}\rho\left(\boldsymbol{x},\boldsymbol{x}_{h_{i}}\right)\hat{\omega}_{h_{i}}}.}\label{eq:huek}
\end{equation}
The integrand on (\ref{eq:huek}) gives a constant which is normalized
to unity. From (\ref{eq:huek}), it is clear that for a case where
curvatures of the manifold are concentrated on hinges (conical singularity),
the holonomy (or deficit angle) does not depend on the area of surface
enclosed by the loop, $\left|\alpha_{S_{i}}\right|$ (not to be confused
with $\left|\alpha_{h_{i}}\right|,$ which is the weight of the infinitesimal
'loop' (plane) orientation). An important fact one needs to notice
is the simplicity of $J_{i}$ as the algebra of $U_{S_{i}}$. This,
at least in 4D, cause the simplicity of $U_{S_{i}}.$ So we could
conclude an important fact: Besides of the simplicity of all bivectors
constructing the simplices, \textit{in 4-dimensional Euclidean Regge
Gravity, all the holonomy circling a single hinge $U_{h_{i}}$ are
simple rotations.}

\subsection{The Angle Relation as Contracted Bianchi Identity}

Let us proceed further to an interesting geometrical fact on a simplicial
complex. Without loosing of generality, let us consider a special
$n$-dimensional simplicial complex known as the $(n+1,1)$-Pachner
move. Let us take a $d$-simplex inside this move, labeled as $\Delta^{\left(d\right)}$,
with $d<n-2.$ $\Delta^{\left(d\right)}$ is shared by minimal three
$(d+1)$-simplices. Let us consider three of them, say $\Delta_{i}^{\left(d+1\right)}$
for $i=1,2,3$. Each two of them, say $\Delta_{i}^{\left(d+1\right)}$
and $\Delta_{j}^{\left(d+1\right)}$, define a $(d+2)$-dimensional
angle, which we label as $\phi_{ij}.$ Thus one has three $(d+2)$-angles
$\left\{ \phi_{ij}\right\} $, $j\neq i$ located on a $\left(d+2\right)$-hinge
$\Delta^{\left(d\right)}$. Moreover, each one of the three sets with
elements $\left\{ \Delta_{i}^{\left(d+1\right)},\Delta_{j}^{\left(d+1\right)},\phi_{ij}\right\} $
belongs to a $(d+2)$-simplex, labeled as $\Delta_{ij}^{\left(d+2\right)}$
. In a recursive manner, two of these $(d+2)$-simplices, say $\Delta_{ij}^{\left(d+2\right)}$
and $\Delta_{ik}^{\left(d+2\right)}$, with $i\neq j\neq k$, define
a $(d+3)$-dimensional angle located on a $\left(d+3\right)$-hinge
$\Delta_{i}^{\left(d+1\right)}$, which we label as $\theta_{jk,i}.$
Remarkably, these three sets of angle $\left\{ \phi_{ij}\right\} $
and $\left\{ \theta_{jk,i}\right\} $ satisfy the dihedral angle relation
(\ref{eq:huff-1}), regardless of the dimension of the simplices \cite{Simone}. 

We argue that the dihedral angle relation can be interpreted as the
contracted Bianchi Identity in a simplicial complex. Let us consider
three $n$-hinges on the $(n+1,1)$-Pachner move. The three loops
circling the hinges $h_{i}$, say $\gamma_{h_{i}},$ could be defined
as the boundaries of faces $S_{i}$ in Voronoi dual lattice \cite{Miller}.
As a consequence to this, three of these loops meets on a point $\mathcal{O}_{p}$.
See FIG 4(a). 
\begin{figure}
\begin{centering}
\includegraphics[scale=0.6]{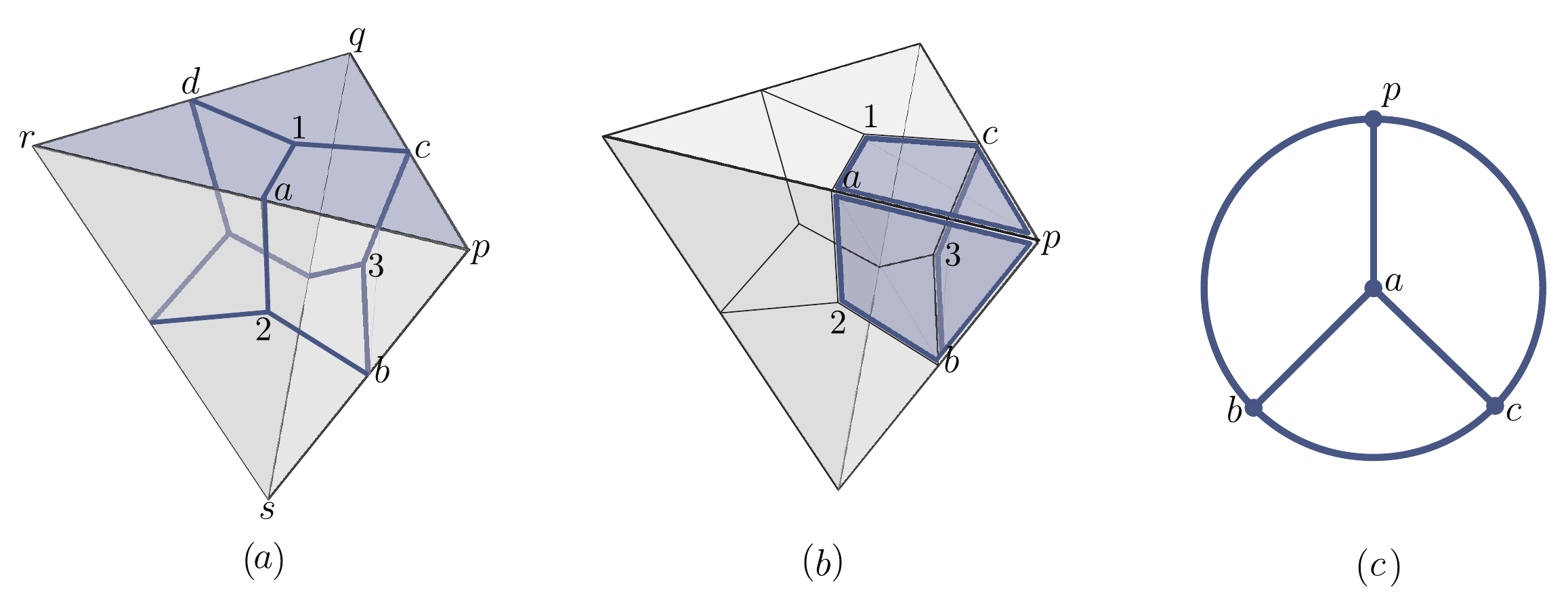}
\par\end{centering}

\caption{(a). The three loops $\gamma_{1}=\gamma_{pq}\gamma_{qr}\gamma_{rp}$,
$\gamma_{2}=\gamma_{pr}\gamma_{rs}\gamma_{sp}$, $\gamma_{3}=\gamma_{ps}\gamma_{sq}\gamma_{qp}$,
meet on point $p$. Inside each 'Voronoi' loop $\gamma_{i}$, lies
the 'Delaunay' $n$-hinges $h_{i}$, which are dual to each other.
'Voronoi' point $p,q,r,s,$ are dual to 'Delaunay' $n$-simplices.
(b) Three subholonomies meeting on $p,$ say $U_{\gamma_{pa1cp}}$,
$U_{\gamma_{pc3bp}}$, $U_{\gamma_{pb2ap}}$. One could product them
together such that $U_{\gamma_{pa1cp}}U_{\gamma_{pc3bp}}U_{\gamma_{pb2ap}}=U_{\gamma_{pa1c3b2ap}}$.
But since the holonomies on internal curves are identity, $U_{\gamma_{pa1c3b2ap}}=1,$
or more general (\ref{eq:bind}). (c) A graph of blue loops configuration
in (b). Loop $\gamma_{a1c3b2a}=\gamma_{acba}$ has identity as its
holonomy, and therefore are equivalent to the theta graph in FIG.
1(a), by an identification $a=b=c=p'$.}
\end{figure}

One could attach elements of group to define holonomies on $\gamma_{h_{i}}$:
\begin{equation}
\left.U_{\gamma_{i}}\right|_{\mathcal{O}_{p}}=\left.U_{S_{i}}\right|_{\mathcal{O}_{p}}=\exp\left.\hat{J}_{i}\right|_{\mathcal{O}_{p}}\delta\theta_{i}\in SO(n).\label{eq:holoholo}
\end{equation}
$\left.\hat{J}_{i}\right|_{\mathcal{O}_{p}}$ is exactly $\hat{\omega}_{h_{i}}'=e^{-1}\hat{\omega}_{h_{i}}e$
as seen from a point $\mathcal{O}_{p}$ inside simplex $p$, where
$\hat{\omega}_{h_{i}}$ is defined as (\ref{eq:hi}). On each hinge
$h_{i}$, the deficit angles are located, satisfying:
\begin{equation}
\delta\theta_{i}=2\pi-\left(\theta_{i,p}+\theta_{i,q}+\theta_{i,r}\right),\label{eq:defsc}
\end{equation}
where $\theta_{i,p}$ is the $n$-dimensional angle of simplex-$p$
located on $n$-hinge $h_{i}.$ One could define a special decomposition
on the holonomy such that:
\begin{equation}
\left.U_{S_{i}}\right|_{\mathcal{O}_{p}}=\underset{\left.U_{i,p}\right|_{\mathcal{O}_{p}}}{\underbrace{\exp-\left.\hat{J}_{i}\right|_{\mathcal{O}_{p}}\theta_{i,p}}}\underset{\left.U_{i,q}\right|_{\mathcal{O}_{p}}}{\underbrace{\exp-\left.\hat{J}_{i}\right|_{\mathcal{O}_{p}}\theta_{i,q}}}\underset{\left.U_{i,r}\right|_{\mathcal{O}_{p}}}{\underbrace{\exp-\left.\hat{J}_{i}\right|_{\mathcal{O}_{p}}\theta_{i,r}}},\quad i=1,2,3,\label{eq:decompose}
\end{equation}
this is illustrated in FIG. 4(a), with the following explanation.
The holonomy on $\gamma_{i}=\partial S_{i}$ is (\ref{eq:holoholo}).
Let us take $\left.U_{\gamma_{1}}\right|_{\mathcal{O}_{p}}$ as an
example. Moving the origin from $p$ to $a,$ such that $\left.U_{\gamma_{1}}\right|_{\mathcal{O}_{a}}=U_{\gamma_{ap}}\left.U_{\gamma_{1}}\right|_{\mathcal{O}_{p}}U_{\gamma_{ap}}^{-1}$,
it is clear that $\left.U_{\gamma_{1}}\right|_{\mathcal{O}_{a}}=U_{\gamma_{apc}}U_{\gamma_{cqd}}U_{\gamma_{dra}}$
which is a product of three holonomies on open curves. One could choose
such that the three subholonomies have $\theta_{1,p},\theta_{1,q},\theta_{1,r}$
from (\ref{eq:defsc}) as their modulus of rotation. Now we want to
decompose $\left.U_{\gamma_{1}}\right|_{\mathcal{O}_{a}}$ such that
it consists a product of holonomies on closed curve as follows: $\left.U_{\gamma_{1}}\right|_{\mathcal{O}_{a}}=U_{\gamma_{apc1a}}U_{\gamma_{a1cqd1a}}U_{\gamma_{a1dra}}.$
Let us choose a special gauge fixing such that the holonomies on internal
curves are identities, say $U_{\gamma_{1a}}=U_{\gamma_{1c}}=U_{\gamma_{1d}}=1$.
Therefore, $U_{\gamma_{apc1a}}=U_{\gamma_{apc}}$, $U_{\gamma_{a1cqd1a}}=U_{\gamma_{cqd}},$
and $U_{\gamma_{a1dra}}=U_{\gamma_{dra}}$. Sending back these holonomies
from $a$ to $p,$ they clearly describe the decomposition defined
in (\ref{eq:decompose}). 

Doing the same decomposition procedure to $\left.U_{\gamma_{2}}\right|_{\mathcal{O}_{p}}$
and $\left.U_{\gamma_{3}}\right|_{\mathcal{O}_{p}},$ one obtains
three subholonomies meeting on $p,$ say $U_{\gamma_{pa1cp}}$, $U_{\gamma_{pc3bp}}$,
$U_{\gamma_{pb2ap}}$, or using more compact notations:
\begin{equation}
\left\{ \left.U_{1,p}\right|_{\mathcal{O}_{p}}=\exp-\left.\hat{J}_{1}\right|_{\mathcal{O}_{p}}\theta_{1,p},\;\left.U_{2,p}\right|_{\mathcal{O}_{p}}=\exp-\left.\hat{J}_{2}\right|_{\mathcal{O}_{p}}\theta_{2,p},\;\left.U_{3,p}\right|_{\mathcal{O}_{p}}=\exp-\left.\hat{J}_{3}\right|_{\mathcal{O}_{p}}\theta_{3,p}\right\} ,\label{eq:in}
\end{equation}
See FIG. 4(b). It is clear that the product of holonomies in (\ref{eq:in})
is equal to identity:
\begin{equation}
\left.U_{1,p}\right|_{\mathcal{O}_{p}}\left.U_{2,p}\right|_{\mathcal{O}_{p}}\left.U_{3,p}\right|_{\mathcal{O}_{p}}=1,\qquad\left.U_{i,p}\right|_{\mathcal{O}_{p}}\in\rho_{n}\left[SU\left(2\right)\right]_{\textrm{sim}}\subset SO\left(n\right).\label{eq:bind}
\end{equation}
By an identification of point $a=b=c$ in FIG. 4(c), the configuration
of loops where $\left.U_{1,p}\right|_{\mathcal{O}_{p}},$ $\left.U_{2,p}\right|_{\mathcal{O}_{p}}$
, and $\left.U_{3,p}\right|_{\mathcal{O}_{p}}$ are attached is topologically
equivalent to a theta graph in FIG 1. Therefore relation (\ref{eq:bind})
is indeed the Bianchi Identity. 

One could realize the following facts that: (1) $\left.U_{i,p}\right|_{\mathcal{O}_{p}}$
are simple rotations since $\left.\hat{J}_{i}\right|_{\mathcal{O}_{p}}$
are simple, and (2) $\left\{ \left.\hat{\omega}_{h_{i}}\right|_{\mathcal{O}_{p}}\right\} $
and thus $\left\{ \left.\hat{J}_{i}\right|_{\mathcal{O}_{p}}\right\} ,$
$i=1,2,3$ construct a trihedron, which cause $\left\{ \left.U_{i,p}\right|_{\mathcal{O}_{p}}\right\} $
$i=1,2,3$ belongs to a common SU(2) (or SO(3)) subgroup of SO(n).
With the Bianchi Identity (\ref{eq:bind}), the set $\left\{ \left.U_{i,p}\right|_{\mathcal{O}_{p}}\right\} $
$i=1,2,3$ satisfies either Theorem 3.4 or 3.2. As a consequence to
this, the trace of (\ref{eq:bind}) in the form of (\ref{eq:3.3-1})
gives angle relation (\ref{eq:huff-1}). For consistency, one could
check whether the angle $\left\{ \theta_{i,p},i=1,2,3\right\} $ really
satisfy the angle relation from the geometries of the $(n+1,1)$-Pachner
move: In fact, the angle $\left\{ \theta_{i,p}\right\} ,$ $i=1,2,3,$
are the angles between $(n-1)$-simplices located on hinges $h_{i}$,
say $\Delta_{i}^{\left(n-2\right)}$, where these hinges meet on a
common $(n-3)$ simplex $\Delta^{\left(n-3\right)}$, and thus needs
to satisfy the dihedral angle relation. With these arguments, \textit{the
dihedral angle relation represents the 'contracted' Bianchi identity
for a simplicial complex}. Moreover, since the simplices satisfies
dihedral angle relation, \textit{the gauge group of discrete gravity
is a simple representation of $SU(2)$, instead of $SO(3)$.}

\section{Discussion and Conclusion}

As we had mentioned in the Introduction, there are three aspects which
becomes our main interest in this article: (1) The gauge group SO(3)
and SU(2), (2) the angle relation, or SU(2) trace relation, or spherical
law of cosine, and (3) the simplicity of the bivectors, and more over
the simplicity of the rotations. By the explanation from the previous
sections, it had been clear that these three properties are related
to one another, nevertheless, we will discuss these relations in more
detailed manner. 

Let us started from the simplicity of bivectors. As already been stated
in \cite{Baez}, in order to construct a simplex from bivectors, each
subsimplices need to be constructed from \textit{simple bivectors
living in the same subspaces}. As we have mentioned earlier, the simplicity
of a bivector guarantees the existence of a single plane defined by
two non-parallel vectors. The existence or these vectors are crucial
for the construction of a simplex, since it is uniquely determined
by its edges. For 4D Regge simplicial complex, the simplicity condition
is equivalent to demand that the norm of the self-dual and anti self-dual
parts of the bivectors are equal. In this article, this is realized
by elements of $\mathfrak{so(4)}$ satisfying Case 1 and Case 3 defined
in Section IV. In fact, in spinfoam model of gravity, one has a more
strict condition concerning the simplicity of the bivector, such that
not only the norm of the self-dual and anti self-dual parts need to
be equal, but also their directions. Precisely, they need to satisfy
only Case 1 of Section IV. The condition is known as the \textit{linear}
simplicity constraint \cite{Carlo1,Immirzi}, which originates from
a specific gauge fixing, the time gauge \cite{karim}. Our work gives
a similar condition for the holonomy representation of discrete gauge
gravity. Through the derivation in Section V A, if we want our (lattice)
gauge theory to describe Regge $n$-dimensional simplicial complex,
each holonomy on the hinges needs to be simple.

The second important aspect is the angle relation. It should be kept
in mind that besides being simple, the bivectors in an $n$-simplex
needs to construct also the lower dimensional subsimplices recursively.
In other words, the bivectors needs to be collected into sub-algebra
space of $\mathfrak{so(n)},$ with $\mathfrak{so(3)}$ being the simplest,
non-trivial case. This becomes the reason why SO(3) and SU(2) become
an interest in this article, in the sense that SO(3) is the 'building
blocks' for higher dimensional orthogonal groups. We have found in
Section V that all holonomies on the hinges are simple rotations,
and moreover one could decompose in a proper way such that a holonomy
on a single hinge is a product of (minimal) three simple subholonomies
meeting on a point, satisfying the Bianchi Identity. Contracting the
Bianchi Identity, one could obtain the trace relation of the holonomy.
Since the subholonomies are simple and they belong to a common SO(3)/SU(2)
group, the trace relation gives either (\ref{eq:3.2}) or the spherical
law of cosine (\ref{eq:realdihed}), by Theorem 3.2 or 3.4. But one
has an interesting fact that the simplex in any dimension always satisfy
dihedral angle relation (\ref{eq:realdihed}). In this step we need
to rule out the SO(3) group, since the one giving the angle relation,
which is a constraint that must be satisfied by a closed Euclidean
complexes, is the spherical law of cosine originating from the trace
relations of SU(2). 

The last aspect is the SO(3)/SU(2) relation. As we had mentioned previously,
SO(3) is the 'building blocks' for higher dimensional orthogonal groups.
The importance of SU(2) is indirect: because it double covers the
SO(3). But we have an interesting fact that the simplex satisfy SU(2)
trace relation instead of SO(3) as its angle relation, and we found
that this is not merely a coincidence. The reason for a simplex to
satisfy SU(2) trace relation might be traced to the fact that an $n$-simplex,
is a special case of convex polytopes, homeomorphic to an $n$-ball,
with the boundary being homeomorphic to an $(n-1)$-sphere \cite{textbook2}.
The trace of the Bianchi Identity carries information about the local
curvature on the boundary. SU(2) is topologically isomorphic to a
3-sphere; a spherical triangle lies on its great 2-sphere. This is
also the reason why SU(2) trace relation gives the spherical law of
cosine. SO(3), although it describe a rotation in 3-dimensional Euclidean
space, is not simply connected as an $\mathbb{RP}^{3}$. These are
also true for higher dimensional rotation: SO(n), in general is not
simply connected. The Spin group, Spin(n), which double covers SO(n),
is usually used as a substitute to SO(n) because of their simpler
topological structure. Nevertheless, in a simplicial complex of Regge
Calculus, the reason of using SU(2) instead of SO(3), is more than
merely a factor of simplicity, but as a natural way which originates
from a property of a simplex as a convex polytopes. In fact, the less
natural feature of SO(3) in describing rotations in 3D, compared to
SU(2), was already known, a common example are the problem of gimbal
lock in navigation \cite{gimbal}, and Dirac belt in a more abstract
way \cite{belt}, with a well-known example in physics includes the
existence of spins in quantum mechanics. The fact that SU(2), defined
in a complex and imaginary manner, provide a more compact and natural
way to handle real-world problems is a fascinating fact of reality. 

To conclude the article, we address the question concerning the holonomy
group for discrete manifold in Regge Calculus. We found that the holonomy
group is restricted such that the holonomies on the loop circling
a single hinge are simple rotations, and that at each points where
these loops meet, the angle relation as the contracted Bianchi Identity,
is satisfied. One of the importance of this result, is that it provide
a discrete and regularized version of the simplicity constraint for
the Lie algebra-valued connection, while other importance are to be
studied elsewhere.

\section*{Acknowledgment}

S. A. is supported by Institut Teknologi Bandung In House Collaboration Post Doctoral Fellowship. F. P. Z.   would like to thank Kemenristekdikti for financial support.

\end{document}